\documentclass[conference]{IEEEtran}

\usepackage{cite}
\usepackage{amsmath,amssymb,amsfonts}
\usepackage{mathtools}
\usepackage{multicol}
\usepackage{algorithm, algpseudocode}
\usepackage{setspace}
\usepackage{enumitem}
\usepackage{breqn}
\usepackage{graphicx}
\usepackage{lipsum}
\usepackage{svg}
\usepackage{textcomp}
\usepackage{xcolor}
\usepackage{booktabs}
\ifCLASSOPTIONcompsoc
\usepackage[caption=false,font=footnotesize,labelfont=sf,textfont=sf]{subfig}
\else
\usepackage[caption=false,font=footnotesize]{subfig}

\def\BibTeX{{\rm B\kern-.05em{\sc i\kern-.025em b}\kern-.08em
    T\kern-.1667em\lower.7ex\hbox{E}\kern-.125emX}}

\begin{document}

\title{A Trust-Based Malicious RSU Detection Mechanism in Edge-Enabled Vehicular Ad Hoc Networks\\
}

\author{\IEEEauthorblockN{Farhana Siddiqua, Mosarrat Jahan}
\IEEEauthorblockA{Department of Computer Science and Engineering, University of Dhaka, Dhaka, Bangladesh\\
Email: 2013-112-145@student.cse.du.ac.bd, mosarratjahan@cse.du.ac.bd}}

\maketitle

\begin{abstract}
Edge-enabled Vehicular Ad Hoc Network (VANET) introduces real-time services and storage, computation, and communication facilities to the vehicles through Roadside Units (RSUs). Nevertheless, RSUs are often easy targets for security assaults due to their placement in an open, unprotected environment and resource-constrained nature. The malicious RSUs compromised by security attacks impose threats to human safety by impeding the operations of VANETs. Hence, an effective malevolent RSU detection mechanism is crucial for VANETs. Existing trust-based detection mechanisms assign trust scores to RSUs based on their interactions with moving vehicles where precise detection of rogue RSUs depends on the accuracy of trust scores. However, brief interaction of RSUs with the running vehicles permits inadequate time to estimate trust accurately. Besides, current works use only vehicle speed and density in beacon messages to assess trust without considering the sensor-detected data in the same messages. Nonetheless, sensor data is useful for traffic management, and neglecting them creates inaccuracy in trust estimation. In this paper, we address these limitations and propose a trust-based scheme to detect malicious RSUs that uses stable and frequent RSU-to-RSU (R2R) interaction to precisely analyze the behavior of an RSU. We also offer a mechanism to detect alteration of sensor-detected data in beacon content and incorporate this scheme in the trust calculation of RSUs. The experimental results show that the proposed solution effectively detects approximately 92\% malicious RSUs, even in the presence of hostile vehicles. Moreover, integrating the proposed solution with the VANET routing protocols improves routing efficiency.
\end{abstract}


\section{Introduction}
Vehicular Ad Hoc Network (VANET) is a leading-edge technology to improve traffic management systems. It models the traffic system as an ad hoc network and enables information exchange among the running vehicles  \cite{onieva2019edge}, \cite{qu2015security}. It enhances road safety, reduces road accidents and traffic jams, and creates provisions for smart travel planning \cite{onieva2019edge}, \cite{sheikh2019survey}. Due to the highly dynamic environment of VANET, the availability of correct information at the right moment is a prime requirement for its proper operation. In this regard, Roadside Units (RSUs) play a vital role in accelerating information processing and providing services at low latency. However, the advantage of minimum latency comes with the cost of numerous security and privacy issues introduced by RSUs, affecting information accuracy \cite{abhishek2019detecting}.

RSUs are edge devices usually deployed along the roadside on traffic lights, bus stops, road signs, etc., to provide various services to the vehicles \cite{onieva2019edge}, \cite{kang2018blockchain}. They can be installed at any place but are cost-effective to deploy in places where traffic volume is high, and placement of expensive general-purpose edge devices with higher computing facilities is not feasible \cite{onieva2019edge}. Although RSUs have enough resources to serve the vehicles in their coverage area, they are typically resource-constrained compared to the general-purpose edge devices \cite{onieva2019edge}. Due to the outdoor placement without tight protections from network operators, RSUs are vulnerable to intrusions, physical attacks, malfunctions, node compromise, sensor tampering attacks, etc. \cite{yang2018blockchain}, \cite{van2018survey}. Moreover, they cannot support computation-intensive security mechanisms due to their resource-constrained nature, making them easy victims of various security attacks. However, RSUs are trusted with several important responsibilities such as vehicle authentication \cite{yao2019bla}, rogue vehicle detection \cite{al2019privacy}, and revocation \cite{malik2018blockchain}. Therefore, RSUs compromised by security attacks severely affect the correct functioning of the entire system, leading to severe consequences threatening human safety. Hence, accurate identification of malicious RSUs and avoiding them from the VANET operation is essential to ensure safe, secure and time-sensitive operation of VANETs.

Traditional cryptography-based security mechanisms are not suitable for dynamic-nature VANET due to their time-consuming and intensive computations and the inability to address some security attacks such as false data injection and internal attacks \cite{zaidi2014data}. Hence, trust-based schemes have been considered as an alternative to identify malicious entities in VANET at a low-cost \cite{hussain2020trust}. In this respect, Abhishek et al. \cite{abhishek2019detecting} addressed the problem of detecting malicious RSUs by evaluating the trust values of RSUs based on Vehicle-to-RSU (V2R) communication, where a vehicle rates an RSU for acquired services. However, the duration of V2R communications is short due to vehicles' high mobility, providing insufficient time to precisely estimate trust values. Hence, inaccuracy in trust estimation allows malicious RSUs to remain undetected and disrupt the functionalities of VANETs silently. Besides providing services to vehicles, RSUs also interact with their one-hop neighbor RSUs for message routing \cite{mershad2012roamer}, periodic beacon message broadcasting \cite{maglaras2013exploiting}, and traffic alert sharing \cite{jindal2017reducing}. They use message routing to forward vehicles' data packets \cite{mershad2012roamer} and utilize beacon messages to share traffic information with other RSUs and vehicles \cite{maglaras2013exploiting}. Besides, RSUs share traffic alerts to avoid unwanted situations such as accidents and bad road conditions \cite{jindal2017reducing}. This indicates that R2R transmissions occupy most of the communications that an RSU uses to contact the remaining entities of the VANET. Hence, the trustworthiness of an RSU should also reflect its reliable behavior in all these aspects, which is missing in the existing literature. In contrast to V2R communication, R2R communications are stable as the positions of RSUs are static, and they frequently interact with their one-hop neighbor RSUs. Therefore, we believe R2R communications can enable precise and error-free trust calculation by allowing an RSU in place of a vehicle to evaluate the trust of other RSUs.
Besides, existing works verify vehicle speed and density to determine the legitimate beacon content, and these parameters are used in the trust calculation based on beacon messages \cite{arshad2018beacon}, \cite{zaidi2015host}. Apart from vehicle speed and density, an RSU also shares sensor-detected data such as humidity \cite{jindal2017reducing}, temperature \cite{jindal2017reducing}, and carbon emission level \cite{maglaras2013exploiting} in beacon messages. The correctness of these sensor-detected data is also crucial as they directly impact traffic management; for example, vehicles usually try to avoid industrial areas prone to excessive carbon emissions. Hence, ignoring sensor-detected data to verify the validity of beacon content can create difficulty in traffic management. Therefore, trust calculation based on the beacon content should also reflect the correctness of sensor data, which is not considered in the literature.

In this paper, we address the shortcomings mentioned above and propose a malicious RSU detection mechanism based on trust calculations that evaluates an RSU's behavior depending on its interaction with other RSUs in the VANET. In particular, we make the following contributions:

\begin{itemize}
\item We propose a trust-based mechanism that assesses an RSU for its behavior in all the R2R communications. Besides, we incorporate an equation to compute an aggregated trust score of an RSU that uniquely combines its score in individual R2R communication. Finally, we compare the trust values with a threshold to identify malicious RSUs.

\item We incorporate an equation to associate packet forwarding ratio with packet modification attack to measure the trust of an RSU based on message routing.


\item We offer a robust mechanism to detect the correctness of the sensor-detected data in a beacon message. We incorporate a new event type known as IGNORE\_RSU for a traffic alert message to accomplish this verification process.

\item We present a novel means to compute the weight of a beacon message based on its correctness on sensor-detected data. Finally, these weights are combined with the outcome based on the vehicle speed and density verification to generate the final trust based on beacon content.

\item We implement the proposed scheme and evaluate the performance through extensive experiments. The results demonstrate the effectiveness of the proposed solution in detecting malicious RSUs, which is approximately $92\%$ in the presence of rogue vehicles. The experimental results also show that the proposed scheme improves the routing efficiency of the existing VANET routing protocols when incorporated with them. 
\end{itemize}

The remainder of the paper is organized as follows. Section \ref{sec:rel_work} reports the related work. Section \ref{sec:proposed_scheme} describes the system model and the comprehensive working procedure of the proposed scheme. Experimental results are discussed in Section \ref{sec:performance}. Finally, Section \ref{sec:conclusion} concludes the paper. 
\section{Related Work}
\label{sec:rel_work}
In the VANET, RSUs are highly vulnerable to internal and external security attacks due to their placement in an unrestricted environment \cite{qu2015security}. Although authentication and cryptography are the first line of defense for VANET, they cannot ensure security against insider attacks \cite{hussain2020trust}. An authenticated entity in the VANET, be it a vehicle or an RSU, may turn into a malicious entity due to defective sensors or node compromise by various attacks \cite{zaidi2014data}. These entities disrupt the secure functioning of VANETs by dropping packets, distorting message content, and injecting false messages \cite{kang2018blockchain}. Therefore, handling these malicious entities is essential for the reliable operation of the VANET. In this respect, trust-based mechanisms play a fundamental role in preventing insider attacks. These schemes assign trust scores to VANET entities based on their behavior to measure their credibilities \cite{hussain2020trust}, \cite{soleymani2015trust}, \cite{guleng2019decentralized}. Although there are numerous works on the trust mechanisms for detecting malicious vehicles \cite{hussain2020trust}, \cite{soleymani2015trust}, \cite{tripathi2019trust}, very few papers consider the issue of identifying the rogue RSUs.

Abhishek et al. \cite{abhishek2019detecting} proposed a trust-based mechanism where every vehicle sends feedback about the RSU it has interacted with to a central trusted server. In this regard, a vehicle evaluates an RSU based on the channel quality and the total number of packets received from or transmitted to the RSU. The central trusted server calculates an aggregated trust value for every RSU based on the received feedback that is later compared with a threshold value to classify the RSU as malicious or authentic.  They also employed a Gaussian kernel-based similarity metric mechanism to handle the impact of false feedback from misbehaving vehicles. However, this model only handles selective packet modification attacks performed by RSUs during V2R communication, which partly reflects the behavior of an RSU. Besides, Lu et al. \cite{lu2018learning} used the physical (PHY)-layer properties such as Received Signal Strength Indicator (RSSI) of the ambient radio signal to detect rogue edge nodes. In this scheme, resource-limited smartwatches and smartphones inside a car outsource heavy computation to an edge node located inside the vehicle. In this case, an outside malicious edge node situated in the VANET environment can launch a man-in-the-middle attack by sending messages to the mobile devices requesting services. The mobile device uses the physical layer properties to distinguish ambient radio signal traces of an outside edge node from an inside legitimate edge node. However, this solution cannot ensure the content accuracy shared by the edge devices, which is crucial for the reliable operation of the VANET.
Moreover, Hao et al. \cite{hao2008distributed} proposed a distributed key management scheme where a trusted authority plays the role of the key generator and an RSU as a key distributor. An RSU forms a group with the vehicles within its transmission range and provides them the group key after confirming their authenticity. After receiving any complaints about other misbehaving vehicles, the trusted authority takes help from the RSU to recover the malicious vehicle's real identity. In this case, a compromised RSU might provide the signature of a legitimate vehicle instead of the malicious one to the authority. The proposed scheme prevents this issue by not providing RSU any access to vehicles' private keys. The main goal of this work is to ensure that RSU performs its duty accurately as a key distributor not to identify the rogue RSU. In contrast to the previous works, our proposed solution analyzes the behavior of an RSU in all the means it can communicate with other RSUs in the VANET and produces an aggregated trust value reflecting realistic conjecture on the reliability of an RSU. Hence, the proposed scheme is resilient to a broad range of security attacks as it assigns trust scores based on the behavior analysis of RSUs.

Existing research works that explore false data detection mechanisms in VANETs mainly consider vehicles' speed and density shared in beacon messages. Existing research exploring false data detection mechanisms in VANETs mainly considers vehicles' speed and density shared in beacon messages. For example, Arshad et al. \cite{arshad2018beacon} proposed a trust management system and fake data detection scheme that utilizes vehicle speed and density shared in beacon messages to measure the trustworthiness of a vehicle. This scheme also uses beacon and safety messages to filter out incorrect messages and get facts from data to assess traffic data reliability. Besides, Zaidi et al. \cite{zaidi2015host} proposed an Intrusion Detection System (IDS) that also uses speed and density collected from neighbor vehicles' beacon messages to identify rogue vehicles. The proposed IDS runs statistical analysis on the collected data to detect false information attacks. In another work, Al-Otaibi et al. \cite{al2019privacy} classified traffic data using vehicle speed to identify rogue vehicles. In this work, an RSU analyzes traffic data provided by vehicles within its transmission area to calculate an estimated speed range. A vehicle is rogue if its speed does not belong to the calculated speed range. Similarly, Paranjothi et al. \cite{paranjothi2020f} also utilized vehicles' speed to detect rogue vehicles. This scheme chooses guard vehicles with a more significant number of neighboring nodes. These vehicles perform a hypothesis test on the neighbor vehicles' speed to identify malicious vehicles. Moreover, Liu et al. \cite{liu2020detecting} proposed a false message detection scheme that uses a traffic flow model to analyze the actual behavior of a traffic environment by utilizing vehicle speed and density collected from neighbor vehicles' beacon messages. Alongside, Pham et al. \cite{pham2018adaptive} proposed a context-aware trust management scheme to evaluate the trustworthiness of the received events by considering the sender vehicles' reputation. This scheme runs an evaluation test on the parameters shared in beacon messages to assess the reliability of an event. In addition, Falasi et al. \cite{al2016similarity} designed a trust management scheme using similarity-based trust relationships to detect false event alerts generated by compromised vehicles. This scheme also uses beacon content such as location and vehicle speed to predict the reliability of a reported event. In contrast to prior works, our scheme considers sensor-detected data as well as speed and density information in the beacon messages to verify the validity of traffic data and incorporates both types of data to evaluate trust based on beacon content.
\section{Proposed Scheme}\label{sec:proposed_scheme}
In this section, we present a trust-based solution to detect malicious RSUs in the VANET. We first discuss the system model for the proposed scheme, followed by a detailed description of the working principle of the proposed scheme.
\subsection{System Model}
\label{sec:sys_mod}
\begin{figure}
\centering
\includegraphics[width=0.45\textwidth]{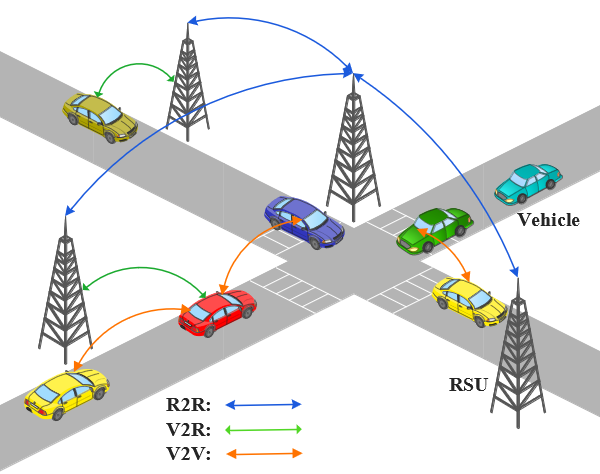}
\caption{System model of the proposed scheme.}
\label{fig:sys_mod}
\end{figure}
Figure \ref{fig:sys_mod} presents the system model of our proposed scheme. It comprises two entities, \textit{vehicles} and \textit{RSUs}. \textit{Vehicles} are equipped with an Onboard Unit (OBU), a Global Positioning System (GPS), and different types of sensors to collect information from their surroundings \cite{onieva2019edge}. They exchange periodic beacon messages, which is also known as Basic Safety Message (BSM) \cite{van2018survey} and Cooperative Awareness Message (CAM) \cite{jin2018expedited} to inform their existence and provide traffic information perceived through their sensors. They also share traffic alerts to notify emergency events to other vehicles and the nearest RSU.
The communication among the vehicles known as Vehicle-to-Vehicle (V2V) communication ranges from $50$ to $300$ meters \cite{zhang2019data}. On the other hand, the communication between an RSU and vehicles is known as V2R communication. \textit{RSUs} are local cloud servers placed at less than the one-kilometer distance in an area where traffic volume is usually high \cite{zhang2019data}. They are equipped with a network device supporting IEEE 802.11p protocol, devices to communicate with the infrastructure network, a GPS, and sensors \cite{van2018survey}. 
They provide real-time services to vehicles and process the data collected from vehicles through V2R communications \cite{al2019privacy}. Besides, RSUs generate beacon messages periodically \cite{maglaras2013exploiting} and share alert messages when emergency events occur \cite{jindal2017reducing}. They also work as a relay node to propagate messages generated by vehicles and RSUs \cite{mershad2012roamer}. An RSU communicates with other RSUs via R2R communication, and the communication range is limited to $1000$ meters \cite{zhang2019data}. We consider a hop count of $6$ to transmit messages generated by RSUs as traffic data can be relevant up to $5km$ \cite{lee2014vehicular} and the maximum distance between two RSUs is $1km$.

\subsection{Overview of the Proposed Scheme}
\label{sec:proposed_scheme_overview}
\begin{figure}
\centering
\includegraphics[width=0.45\textwidth]{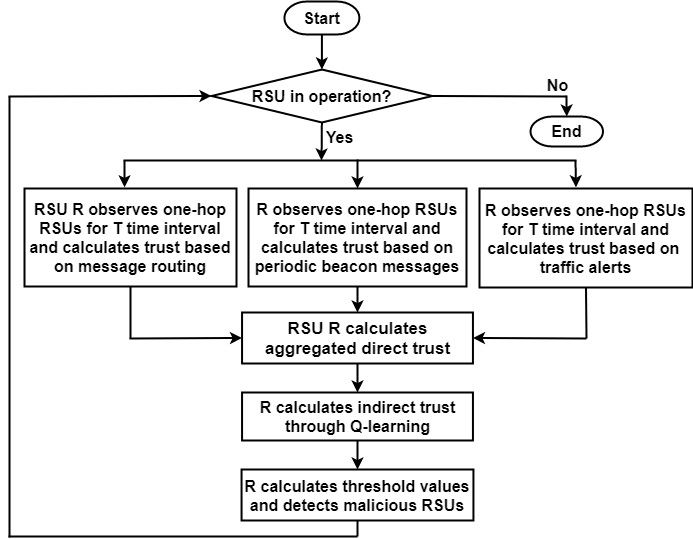}
\caption{Flow diagram of the proposed scheme for RSU $R$.}
\label{fig:RFlowDiagram}
\end{figure}

Each RSU $R$ monitors its one-hop neighbor RSUs for a pre-defined time duration $T$ shown in Fig. \ref{fig:RFlowDiagram}. During this time, $R$ observes the behavior of its one-hop neighbor RSUs for routing messages, broadcasting beacon messages, and transferring traffic alerts. When $T$ expires, $R$ assigns trust scores to its neighbors in each of the above-mentioned communication scenarios based on their behavior and eventually combines all the trust scores to calculate the direct trust of the one-hop neighbor RSUs. For non-one hop RSUs, $R$ uses the Q-learning mechanism \cite{guleng2019decentralized} to determine their trust values. $R$ also decides a threshold value for every other RSUs in the network and compares it with the corresponding trust value to identify an RSU as \textit{legitimate} or \textit{compromised} node. 

\subsection{Trust Calculation based on Message Routing}
The proposed scheme observes the behavior of an RSU for dropping packets and altering packet content while it acts as a relay node to route messages from a source vehicle to a destination vehicle \cite{mershad2012roamer}, or forward packets from remote RSUs to a central RSU \cite{huang2017efficient}. Each RSU $R_1$ forwards packets to its one-hop neighbor RSU $R_2$ and monitors the behavior of $R_2$ on forwarding these packets for a fixed time interval $T$. $R_1$ counts the number of forwarded and dropped packets by $R_2$ and checks whether the forwarded messages are maliciously modified. Based on the observation, $R_1$ calculates the trust value,  $Trust_{routing}$ of $R_2$ as follows:
\begin{equation}
\centering
    Trust_{routing} = \frac{PF}{PF+PD} \times PM
\end{equation}	
where $PF$ is the number of packets forwarded, $PD$ is the number of packets dropped, and $PM$ is the packet modification parameter. 

Each RSU $R_1$ uses a Watchdog module \cite{marti2000mitigating} that overhears the incoming and outgoing traffic of other entities within $R_1$'s transmission range. Hence, a watchdog module of $R_1$ can detect whether $R_2$ forwards a packet towards the next node. This module stores the recently sent packets by $R_1$ and removes a packet from the buffer when it overhears the same packet being forwarded by the next-hop RSU $R_2$. In this case, $R_1$ increments $PF$. If a packet remains in the buffer for more than the expected time $t_{expected(R_1, R_2)}$, $R_1$ considers $R_2$ has dropped that packet and increments $PD$. $R_1$ calculates $t_{expected(R_1,R_2)}$ as follows \cite{bhoi2014ijs}:
\begin{equation}
\label{eqn:texpect}
    t_{expected(R_1,R_2)} = \frac{L}{r_{(R_1,R_2)}} + \frac{d_{(R_1,R_2)}}{V_{propagation}} + t_{other}
\end{equation}	
where $L$ is the length of message, $r_{(R_1, R_2)}$ is the data transmission rate, $d_{(R_1,R_2)}$ is the distance between $R_1$ and $R_2$, $V_{propagation}$ is the propagation speed, and $t_{other}$ represents the queuing and processing delay.

The watchdog mechanism also compares the hash value of a packet on the incoming interface of the observed RSU with the hash value of the same packet on the outgoing interface \cite{patil2014detecting}. The hash values are computed on the packet fields that are not supposed to change during routing \cite{patil2014detecting}. If both hash values are the same, then no packet modification is performed by next-hop RSU. If watchdog module detects a packet modification then $R_1$ sets $PM=0$, otherwise $PM=1$. 

\subsection{Trust Calculation based on Beacon Messages} \label{sec:trust_cal_beacon_msg}
Each RSU periodically transmits \textit{hello} messages that are beacon messages to its one-hop neighbor RSUs and vehicles to inform its existence and traffic-related information \cite{maglaras2013exploiting}. The proposed scheme considers the beacon message generation rate and the accuracy of beacon message content to detect malicious behavior of RSUs. 

\subsubsection{Trust Calculation based on Beacon Message Generation Rate}
A malicious RSU prevents the message propagation by other entities of the VANET by flooding the communication channel with beacon messages. We use the flooding attack detection mechanism proposed in \cite{sajjad2015neighbor} to detect RSU's misbehavior. An RSU $R_1$ observes its one-hop RSU $R_2$ in a time slot $i$ of length $T$ and counts $B_i(R_2)$, the number of beacon messages generated by $R_2$ during this interval. $R_1$ also keeps track of the beacon message rate of $R_2$ for the latest $Z$ time slots and calculates the weighted average of beacon message generation rate as $B_{avg}(R_2) = \sum\limits_{t=1}^Z (t/Z) \times B_t(R_2)$. If $B_i(R_2) > B_{avg}(R_2)$ in a time slot $i$, then flooding attack is detected, and $R_1$ sets $Trust_{beacon}$ to $0$; otherwise $Trust_{beacon}=1$.

\subsubsection{Verification of Beacon's Content}
Each RSU periodically shares beacon messages with its adjacent RSUs and vehicles within its transmission range to notify its existence, traffic condition, weather forecast, etc. \cite{jindal2017reducing}. Besides, vehicles also share beacon messages with their neighbor vehicles and the nearest RSU. Alongside, an RSU collects data about traffic situations, weather, etc. through its sensors \cite{sheikh2019survey}. Thus, an RSU receives huge volume of data from the beacon messages provided by both vehicles and neighbor RSUs and from its sensors \cite{hussain2020trust}. The RSU analyzes those data, generates aggregated results for different purposes, and includes the analyzed result in the beacon messages to provide a traffic overview to the vehicles and one-hop neighbor RSUs. An RSU usually shares speed \cite{al2019privacy}, density \cite{arshad2018beacon}, temperature \cite{jindal2017reducing}, humidity \cite{jindal2017reducing} and carbon emission level \cite{maglaras2013exploiting} in the beacon messages. If flooding attack is not detected, $R_1$ verifies the content of the $i$-th beacon message received from $R_2$ in two ways. They are:
\begin{itemize}
\item $R_1$ estimates the speed and density of the vehicles coming from the area of $R_2$ \cite{al2019privacy}, \cite{arshad2018beacon} and matches them with the same information in $R_2$'s beacon message. If they vary by a certain threshold value, $T_{H_1}$, $R_1$ sets $i$-th beacon message, $Beacon_{i}=1$; Otherwise, $R_1$ rejects the beacon message by setting $Beacon_{i}=0$.
\begin{figure}
\centering
\includegraphics[width=0.45\textwidth]{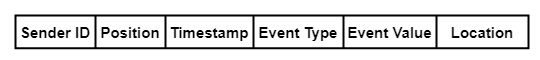}
\caption{Vehicle and RSU traffic alert format.}
\label{fig:VR_alert}
\end{figure}
\item $R_1$ counts the feedback of intermediate vehicles to verify the remaining data in $R_2$'s $i$-th beacon message. A vehicle $V$ in the transmission range of $R_2$ verifies temperature, humidity, and carbon emission level shared in $R_2$'s beacon message through its sensors. If the discrepancy of information in $R_2$'s beacon message and $V$'s sensor data exceeds a threshold value, $T_{H_2}$, $V$ generates an \textit{IGNORE\_RSU} traffic alert indicating invalid content. The neighbor vehicles of $V$ verify the generated alert as they have also received the same beacon message from $R_2$. They share $V$'s alert with their neighbor vehicles if the alert is correct. Otherwise, they discard the traffic alert. Thus, the alert propagates through the VANET and ultimately reaches the one-hop neighbor RSU $R_1$. $R_1$ receives alerts from multiple adjacent vehicles and counts those alerts whose timestamp difference with $R_2$'s $i$-th beacon is less than or equal to a pre-defined threshold value $T_{H_3}$. If the majority of neighbor vehicles agree that $R_2$'s beacon content is inaccurate, $R_1$ sets $Beacon_{i}=0$, otherwise $Beacon_{i}=1$. If $R_1$ does not receive any alert, it sets  $Beacon_{i}=1$. Here to mention that $R_1$ usually verifies speed, density, and sensor data one by one and considers neighbor RSU $R_2$ legal if all of these parameters are accurate. However, $R_2$'s beacon content is considered invalid if $R_1$ finds modification in any of these parameters and does not proceed to check another one.
\end{itemize} 

In our scheme, both vehicles and RSUs generate traffic alerts following the format shown in Fig. \ref{fig:VR_alert} \cite{arshad2018beacon}. It consists of several fields where \textit{Sender ID} is a unique ID of a vehicle or RSU, \textit{Position} is the current position of a vehicle or RSU, \textit{Timestamp} is the traffic alert creation time, \textit{Event Type} indicates traffic event, \textit{Event Value} contains a binary value to indicate the presence and absence of the event and \textit{Location} denotes the event place. We define a new event type \textit{IGNORE\_RSU} to detect invalid data. A vehicle assigns \textit{Event Value}=$0$ for \textit{IGNORE\_RSU} to indicate an RSU at \textit{Location} is a malicious RSU generating false data and $1$ to confirm it as an honest RSU. 


\subsubsection{Trust Calculation based on Beacon Content}
Suppose $R_1$ receives $n$ beacon messages from $R_2$ during $T$. $R_1$ receives the \textit{IGNORE\_RSU} alerts from the adjacent vehicles and determines $Beacon_i$ based on majority vehicles' opinion or the result of its verification of speed and density. $R_1$ calculates a weight $w_i$ for each beacon message $i$ as follows: 
\begin{equation}
\label{eqn:vehicle_weight}
w_{i} = \frac{X_{i}}{\sum\limits_{i=1}^n X_{i}}  
\end{equation}
where $X_{i}$ is the total number of one-hop neighbor vehicles of $R_1$ reporting \textit{IGNORE\_RSU} alert for $i$-th beacon message. If no alert is received for $i$-th beacon message, $X_{i}$ is  the total number of vehicles adjacent to $R_1$. The trust value based on beacon message is calculated as the weighted average of $Beacon_{i}$ generated during $T$ as follows:
\begin{equation}
\label{eqn:beacon_trust}
Trust_{beacon} = \sum\limits_{i=1}^n (w_{i} \times Beacon_{i})
\end{equation}

\begin{figure}
\centering
\includegraphics[width=0.45\textwidth]{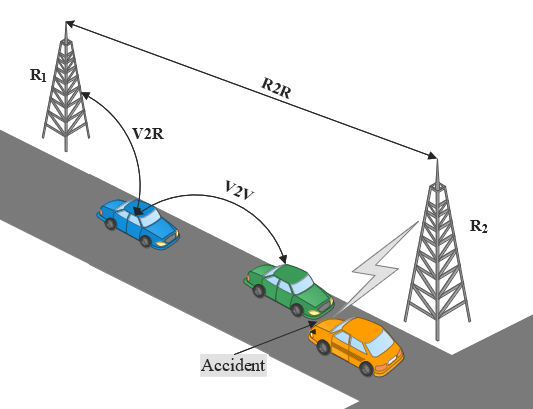}
\caption{R2R false alert detection.}
\label{fig:R2RFalseAlert}
\end{figure}

\subsection{Trust Calculation based on Traffic Alerts} \label{sec:trust_cal_traffic_alert}
Traffic alerts are time-sensitive, and erroneous traffic alerts can cause serious problems such as difficulties in managing traffic and creates threats to human safety \cite{zaidi2014data}. When an emergency event occurs, vehicles observing that event generate and broadcast traffic alerts to the vehicles within their transmission range and the nearest RSU \cite{arshad2018beacon}. Similarly, RSUs noticing the same event also generate and broadcast the traffic alerts to the one-hop neighbor RSUs and vehicles under their coverage \cite{ahmed2018vansec}. A malicious RSU can modify the traffic alert with malicious intent. Thus, the trust of an RSU should also reflect the degree of RSU's capability of transmitting authentic traffic signals.
\begin{algorithm}[H]
\caption{Trust calculation of $R_2$ based on traffic alerts.}
\label{algo:trust_traffic_alert}
\flushleft
\begin{algorithmic}
\Require
\Statex $M_{R_2}:$ traffic alert received from one-hop RSU $R_2$
\Statex $M_V[n]:$ traffic alert received from $n$ adjacent vehicles
\Statex $T_H:$ threshold value for timestamp differences
\State

\State $i = 0$
    \State $Y = N = 0$
    \While{$i<n$}
        \If{$M_{R_2}.Event\_Type==M_V[i].Event\_Type$ and $M_{R_2}.Event\_Value==M_V[i].Event\_Value$ and $M_{R_2}.Location ==M_V[i].Location$ and $M_{R_2}.Timestamp - M_V[i].Timestamp\leq T_H$}
            \State $Y++$
        \Else
            \State $N++$
        \EndIf
        \State $i++$
    \EndWhile
    \State
    \If{$Y>N$}
        \State $Trust_{alert}$=$1$
    \Else
        \State $Trust_{alert}$=$0$
    \EndIf
\end{algorithmic}
\end{algorithm}

Figure \ref{fig:R2RFalseAlert} presents a scenario of how traffic alert is verified in the proposed scheme. Here, RSU $R_2$ monitors the area where an accident took place and RSU $R_1$ is a one-hop neighbor of $R_2$ observing the behavior of $R_2$ for $T$ time duration. Vehicles observing the accident notify $R_2$ about the event and broadcast the traffic alert to vehicles within their transmission range. Besides, $R_2$ also broadcasts a traffic alert for the same accident to $R_1$ and vehicles under its transmission range. Thus, a vehicle receives traffic alerts on the same event from neighbor vehicles and RSUs. It verifies the authenticity of the received traffic alerts using the information sensed by itself or the same traffic alerts received from other neighbor vehicles \cite{guleng2019decentralized} and RSUs. A non-source vehicle considers the content of the maximum number of received alerts on the same event as valid and shares that message with its neighbors. Thus, the alert message propagates from one vehicle to another and ultimately reaches $R_1$. RSU $R_1$ receives traffic alerts on an event from $R_2$ and multiple adjacent vehicles $V$. It calculates the trust value of $R_2$ following Algorithm \ref{algo:trust_traffic_alert}. For each received traffic alert $M_V[i]$ from an adjacent vehicle, $R_1$ compares it with the traffic alert $M_{R_2}$ received from $R_2$. If both messages match on (1) Event Type, (2) Event Value, (3) Location (as shown in Fig. \ref{fig:VR_alert}) and the difference of timestamps is less than or equal to a pre-defined threshold $T_{H}$, $R_1$ considers them as a match and increments $Y$. Otherwise, $N$ is updated. If the number of match, $Y$ is greater than the number of nonmatch, $N$, then $R_1$ sets the trust of $R_2$ based on traffic alert, $Trust_{alert}$ to $1$; Otherwise  $Trust_{alert} = 0$, indicating $R_2$ as a malicious RSU. We can accomplish the same traffic alert verification process using the Decentralized Environmental Notification Message (DENM) \cite{santa2014experimental} in Intelligent Transportation System (ITS).


\subsection{Aggregated Direct Trust Calculation}
$R_1$ calculates the direct trust, $Trust_{direct}$ of $R_2$ as follows: 
\begin{multline}
\label{eqn:dTrust}
Trust_{direct} = (w_1 \times Trust_{routing} + w_2 \times Trust_{beacon})\\ 
 \times Trust_{alert}
\end{multline}
where $Trust_{routing}$, $Trust_{beacon}$, and $Trust_{alert}$ are trust values based on message routing, beacon message broadcasting, and traffic alert sharing, respectively. Besides, $w_1$ and $w_2$ are weights that sum to $1$ and are defined as follows:
\begin{equation}
\label{eqn:wRoute}  
    w_1 = \frac{F_{routing}}{F_{routing} + F_{beacon}}
\end{equation}	
\begin{equation}
\label{eqn:wBeacon}  
    w_2 = \frac{F_{beacon}}{F_{routing} + F_{beacon}}
\end{equation}
where $F_{routing}$ is the frequency of messages for routing, and $F_{beacon}$ is the frequency of beacon messages.

We assign the highest priority to traffic alerts as they are time-sensitive and have severe consequences if they are not delivered in timely order or contain wrong information. Hence, we multiply the weighted sum of $trust_{routing}$ and $trust_{beacon}$ with $Trust_{alert}$ in Eq. \ref{eqn:dTrust} to reflect its precedence.  

We observe that a trust factor with a greater weight can hide the malicious property presented by the other trust factor in the following circumstances:\\
\noindent{1)} $Trust_{beacon}<0.5$ and $w_1 \geq w_2$: The packet routing rate of one-hop RSU is greater than or equal to the beacon message generation rate, and the one-hop RSU behaves maliciously for beacon messages. In this case, $Trust_{routing}$ hides the effect of $Trust_{beacon}$.\\
\noindent{2)} $Trust_{routing}<0.5$ and $w_2 \geq w_1$: $Trust_{routing}$ indicates that the one-hop RSU has dropped more than $50\%$ \cite{xia2018towards} of the packets or it has modified the messages before routing. $w_2 \geq w_1$ indicates that the beacon message generation rate of one-hop RSU is greater than or equal to the packet routing rate. Thus, $Trust_{beacon}$ hides the packet drop attribute of one-hop RSU.  

To handle the above-mentioned cases, we cut off the weight from a trust factor which hides the malicious activities displayed by other trust factor as follows \cite{xie2020novel}:
\begin{equation}
\label{eqn:weight}  
    w = aTe^{-(bT)}
\end{equation}

where $T$ is the pre-defined time interval, $b$ = $w_1$ if $w_1 \geq w_2$ else $w_2$, and $a$ = $1-b$. We assign the new weight $w$ to the trust factor that hides the malicious activities of another trust factor, and $1-w$ is assigned to the remaining trust factor.

\subsection{Indirect Trust Calculation}
We use the Q-learning method \cite{guleng2019decentralized} to compute indirect trust. In this technique, every RSU maintains a Q-table containing an entry for every other RSUs and broadcasts the table with the \textit{hello} messages. In the Q-table, each entry contains $[Q(m,n), X]$, where $Q(m,n)$ is the trust value of $RSU_m$ determined by $RSU_n$, and $X$ is a boolean variable which indicates whether $RSU_m$ is a  one-hop neighbor of $RSU_n$ or not. The initial Q-value of each RSU is $0$. Suppose $RSU_p$ updates the Q-table and the Q-value of $RSU_m$ at $RSU_p$ is updated as follows:

\begin{multline}
Q_{new}(m,p)={\alpha}{\times}Q_{old}(m,p){\times}\{r+{\gamma}{\times}avg_{v{\in}NB_m}Q(m,v)\} \\
+(1-{\alpha}){\times}Q_{old}(m,p)
\end{multline}
where $Q_{new}(m,p)$ is the new trust value of $RSU_m$ evaluated by $RSU_p$, $Q_{old}(m,p)$ is the previously assigned trust value, $NB_m$ is the set of one-hop neighbor RSUs of $RSU_m$, $\alpha$ is the learning rate set to $0.7$ \cite{guleng2019decentralized}, $r$ = $Trust_{direct}$ if $RSU_m$ is a one-hop neighbour of $RSU_p$; otherwise, $r=0$, and $\gamma$ is the discount factor set to $0.9$ \cite{guleng2019decentralized}. The equation takes the average of the trusts provided by the neighbor RSUs to handle the Q-table modification by malicious RSUs and it is denoted as $avg_{v{\in}NB_m}Q(m,v)$.

Significant communication overhead occurs due to huge message passing to update the entries of Q-table for each RSU in the network. To minimize this communication overhead, we limit both the Q-table size and the number of times the Q-table should be broadcasted to $6$ following the hop count constraint discussed in section \ref{sec:sys_mod}.

\subsection{Threshold Calculation and Malicious RSU Detection}
Each RSU exhibits different behavior from the others. Therefore, it is crucial to maintain an individual threshold value for every RSU. The proposed scheme uses the threshold adjustment mechanism \cite{kerrache2018uav} to identify malicious RSUs where the initial trust value and threshold value for each RSU are set to $0.5$. The threshold value varies in the range of $[0.5, 1]$, and the trust value varies in the range of $[0, 1]$. The threshold value is adjusted according to the changes in trust value which reflects the behavior changes of an RSU. The new threshold value, $TH_{new}$ is determined as follows:
\begin{equation}
\label{eq:threshold_generation}
  TH_{new} =
    \begin{cases}
      \beta+$0.5$ & \text{if $\beta>0$}\\
      TH_{old}  & \text{if $\beta=0$}\\
      $0.5$ & \text{if $\beta<0$}
    \end{cases}       
\end{equation}
where $\beta$ is expressed as $\beta=Trust_{old}-Trust_{new}$. 

Each RSU computes trust values of other RSUs in the network and compares the trust value with the corresponding threshold value to identify them as \textit{legitimate} or \textit{compromised} RSU. An RSU is classified as \textit{legitimate} when trust value is higher than $TH_{new}$. Otherwise, it is a \textit{compromised} RSU.

\subsection{Handling Malicious Vehicles}
The proposed scheme uses intermediary vehicles between the one-hop neighbor RSUs to verify beacon content and traffic alerts. Although it is not possible to completely overcome the impact of malicious vehicles, we incorporate a mechanism to minimize it. A malicious vehicle in the transmission path can create a false traffic alert or drop or modify the correct alert messages. In our scheme, nonmalicious neighbor vehicles observing the same event verify the traffic alert's authenticity and drops the message if it is incorrect. They share the message with the vehicles within their transmission range, only if it is valid. Vehicles not observing the event rely on the majority opinions of neighbor vehicles and source RSU and verify the alert accordingly. To reduce the effect of malicious vehicles further, the RSU prefers the opinion of the majority of its adjacent vehicles regarding any event message. Section \ref{sec:trust_cal_beacon_msg} and \ref{sec:trust_cal_traffic_alert} discuss the handling of rogue vehicles in details for verification of beacon messages and traffic alerts, respectively.
\section{Simulation Results}
\label{sec:performance}
In this section, we discuss the performance of the proposed scheme based on several experiments. Besides, we also analyze the impact of the proposed method on the current VANET routing protocols.

\subsection{Experimental Setup}\label{sec:exp_setup}
\begin{figure}
\centering
    \includegraphics[width=0.45\textwidth]{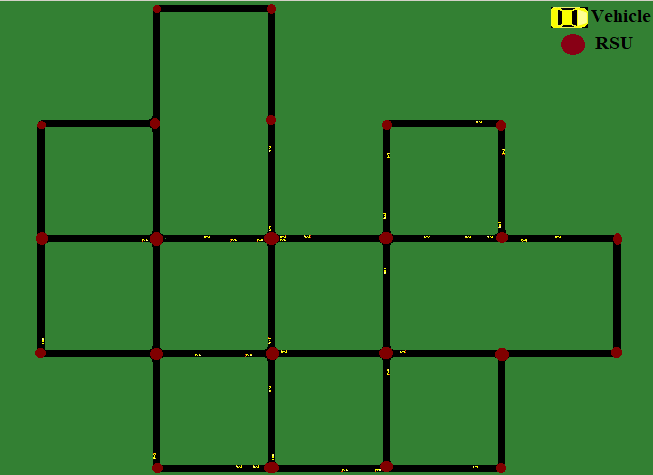}
\caption{Simulation traffic model.}
\label{fig:SimModelVehicles}
\end{figure}
We used Network Simulator 3 (NS-3) \cite{riley2010ns} to evaluate the performance of our proposed scheme. Besides, we used Simulation of Urban Mobility (SUMO) \cite{behrisch2011sumo} to construct a realistic vehicular mobility model. A visual network editor of SUMO known as NetEdit \cite{behrisch2011sumo} was used to insert and manually position static RSU nodes at every intersection in the traffic environment. After that, vehicles were loaded into this network model. Moreover, SUMO configuration files were used to generate trace files containing the information of vehicle movements, and they were fed into the NS-3 simulation. Figure \ref{fig:SimModelVehicles} presents our simulation model generated by SUMO.

For simulation, we considered an area of $14 km \times 14 km$ where $25$ RSUs were serving $500$ vehicles running with an average speed of $20m/s$. NS-3 uses IEEE 802.11p communication protocol, which sets the transmission range of vehicles and RSUs to $250m$ and $900m$, respectively. A summary of the simulation parameters is presented in Table \ref{tab:tab_parameter}.
\begin{table} 
\centering
\caption{Parameters used in Simulation}\vspace{-3mm}
\begin{tabular}{l|l}\specialrule{.12em}{1em}{0em}
\hline
\textbf{Parameter}                 & \textbf{Value}     \\ \hline
Simulation area            & $14 km \times 14 km$       \\ \hline
Number of vehicles         & $500$                      \\ \hline   
Transmission range of vehicle & $250m$                  \\ \hline
Vehicle speed      & $20m/s$                            \\ \hline
Number of RSUs              & $25$                      \\ \hline 
Distance between RSUs      & $900m$                     \\ \hline
Transmission range of RSU & $900m$                      \\ \hline
Simulation time      & $20m$                   \\ \hline
\specialrule{.12em}{0em}{0em}
\end{tabular}
\label{tab:tab_parameter}
\end{table}

We considered the impact of both malicious vehicles and malicious RSUs to evaluate the performance, as the verification of both beacon messages and traffic alerts involves vehicles running on the road. Experiments were conducted with an increasing percentage of malicious RSUs, $MR$ ($20\%$, $40\%$, and $60\%$) and an increasing percentage of malicious vehicles, $MV$ ($5\%$, $10\%$, $15\%$, and $20\%$). Each experiment ran the simulation for $20$ \textit{minutes} and the results were averaged over $10$ iterations with error bars indicate $95\%$ conﬁdence intervals \cite{guleng2019decentralized}. In the experiments, malicious RSUs dropped and forwarded packets with a probability of $0.5$. During packet transmission, they modified packets with a probability of $0.5$. Rogue RSUs also created flooding attacks and altered information in beacon messages with a probability of $0.5$. Further, they altered traffic alerts with a probability of $0.5$. On the other hand, malicious vehicles generated \textit{IGNORE\_RSU} alert for an accurate beacon message with a probability of $0.5$. They dropped \textit{IGNORE\_RSU} event or modified the event value with a probability of $0.5$. Besides, they modified the traffic alerts with a probability of $0.5$.
\begin{table}
\centering
\caption{Parameters used in Performance Metrics}  \vspace{-3mm}
\scalebox{0.95}{
\begin{tabular}{p{3.1cm}|p{5cm}}
\specialrule{.12em}{1em}{0em} 
{\textbf {Parameter}} &  {\textbf{Description}} \\ 
\specialrule{.12em}{0em}{0em}

True Positive (TP) &  No. of malicious RSUs identified correctly.\\ \hline
False Positive (FP) & No. of legitimate RSUs identified as malicious RSUs.\\ \hline
True Negative (TN) & No. of legitimate RSUs identified correctly. \\ \hline
False Negative (FN) &  No. of malicious RSUs identified as legitimate RSUs.\\
\hline

\specialrule{.12em}{0em}{0em}
\end{tabular}}
\label{table:metric_parameter}
\end{table}

\subsection{Performance Metrics}
We evaluated the performance of the proposed scheme using five performance metrics. Table \ref{table:metric_parameter} enlists the parameters used to define these performance metrics. 

\noindent\vspace{0.2mm} \textit{1) False Positive Rate (FPR):} It shows the probability of legitimate RSUs to be identified as malicious RSUs as follows: 
\begin{equation}
\label{eqn:fprate}
    False\; Positive\; Rate = \frac{FP}{FP+TN}    
\end{equation}

\noindent\vspace{0.2mm} \textit{2) False Negative Rate (FNR):} It shows the probability of malicious RSUs to be identified as legitimate RSUs as follows: 
\begin{equation}
\label{eqn:fnrate}
   False\; Negative\; Rate = \frac{FN}{FN+TP}
\end{equation}

\noindent\vspace{0.2mm} \textit{3) Precision:} It is the ratio of correctly detected malicious RSUs to the total number of RSUs that are identified as malicious and defined as follows: 
\begin{equation}
\label{eqn:precision}
    Precision = \frac{TP}{TP+FP}
\end{equation}

\noindent\vspace{0.2mm} \textit{4) Recall:} It is the ratio of correctly identified malicious RSUs to the total number of actual malicious RSUs and defined as follows:   
\begin{equation}
\label{eqn:recall}
    Recall = \frac{TP}{TP+FN}
\end{equation}

\noindent\vspace{0.2mm} \textit{5) Accuracy:} It is the ratio of correctly identified malicious RSUs and legal RSUs to the total number of RSUs present in the network and defined as follows:  
\begin{equation}
\label{eqn:accuracy}
    Accuracy = \frac{TP+TN}{TP+FP+FN+TN}
\end{equation}


\subsection{Performance Analysis}
In this section, we present our findings on the performance of the proposed scheme derived from the analysis of different experiments. Although Abhishek's work \cite{abhishek2019detecting} calculates the trust of RSUs, it is not possible to compare this scheme with our proposed method in a meaningful way as each scheme calculates the trust values of RSUs considering the behavior of RSUs in diverse communication scenarios. Hence, we present the results for the proposed system only.

\begin{figure*}
    \centering
    \subfloat[False positive rate]{\includegraphics[ width=0.27\textwidth]{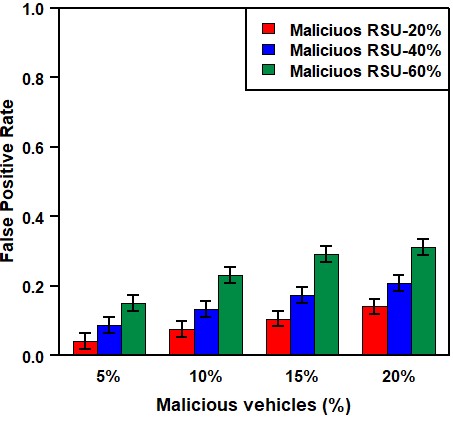}
    \label{fig:fprate}}
    \subfloat[False negative rate]{\includegraphics[ width=0.27\textwidth]{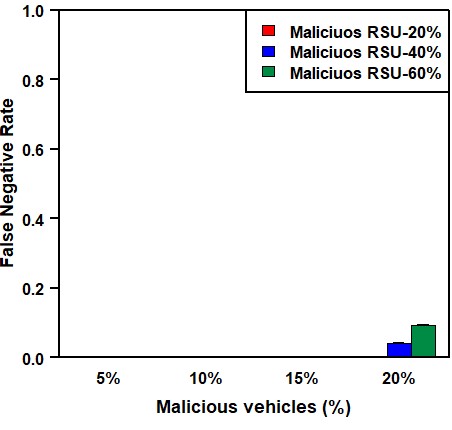}
    \label{fig:fnrate}}
    \subfloat[Precision]{\includegraphics[ width=0.27\textwidth]{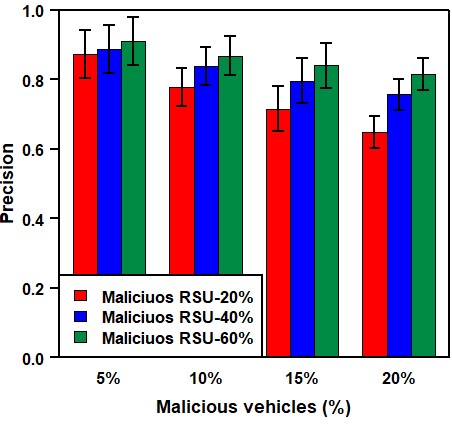}
    \label{fig:precision}}
    \hfil
    \subfloat[Recall]{\includegraphics[ width=0.27\textwidth]{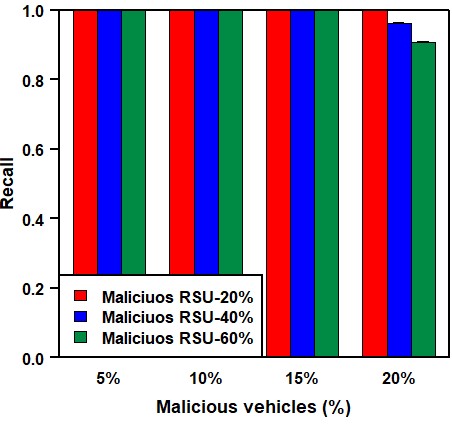}
    \label{fig:recall}}
    \subfloat[Accuracy]{\includegraphics[ width=0.27\textwidth]{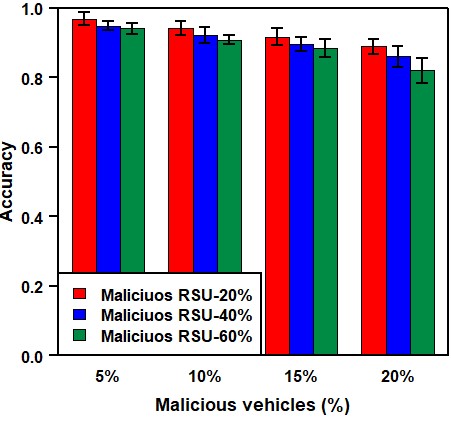}
    \label{fig:accuracy}}
    \hfil
    \caption{Performance analysis of the proposed scheme.}
    \label{fig:performane_analysis}
\end{figure*}
\subsubsection{False Positive Rate (FPR)}\label{sec:FPR}
Figure \ref{fig:fprate} shows that FPR increases with $MV$ and $MR$ and it  reaches to approximately $30\%$ when $MR$=$60\%$ and $MV$=$20\%$. Two reasons are mainly working behind the generation of FPR in our scheme. If $MR$ increases significantly, all the adjacent RSUs of a legal RSU may behave maliciously, and they can bound the legitimate RSU to drop packets by creating beacon flooding attacks. Thus, a legal RSU is identified as a malicious RSU. Besides, vehicles propagate traffic alerts to help verification of beacon content and alert messages. Rogue vehicles may generate \textit{IGNORE\_RSU} alerts for an honest RSU to prove it malicious. Similarly, when an honest RSU reports an event, hostile vehicles can drop or alter the traffic alerts generated by the honest vehicles for the same event and establish the RSU as a malicious one.

\subsubsection{False Negative Rate (FNR)} \label{sec:FNR}
As shown in Fig. \ref{fig:fnrate}, malicious vehicles have very little influence on the FNR. When a malicious RSU generates a false beacon message, the rogue vehicles can either stop propagating \textit{IGNORE\_RSU} or modify the event value. Hence, the malicious attributes of an RSU remain unknown, and it is considered a legitimate RSU. When $MV$ is relatively low, the nonmalicious vehicles suppress malicious vehicles' effect by verifying the same beacon messages. Therefore, FNR is not visible up to $MV$=$15\%$ in Fig. \ref{fig:fnrate}. In our simulation, a situation where modified traffic alerts by malicious RSUs match with the rogue vehicles' opinion does not occur because of the low frequency of traffic alerts and independent decision-making of malicious vehicles without considering the action of malicious RSUs. Therefore, FNR is not generated for traffic alerts. Malicious RSUs have no impact on the FNR. The proposed scheme generates maximum $8\%$ FNR for $MR$=$60\%$ and $MV$=$20\%$ shown in Fig. \ref{fig:fnrate}.

\subsubsection{Precision} 
Figure \ref{fig:precision} shows that the precision decreases with the rising values of $MV$. As we described in Section \ref{sec:FPR}, the impact of $MV$ increases FPR. Thus, the precision decreases with the increasing values of $MV$. For a fixed $MV$, higher values of $MR$ increase both $TP$ and $FP$. Hence, precision increases with $MR$ for a fixed $MV$. The precision reaches almost $81\%$ when $MR$=$60\%$ and $MV$=$20\%$ shown in Fig. \ref{fig:precision}. When $MR$=$60\%$, the precision drops approximately $10\%$ at $MV$=$20\%$ compared with $MV$=$5\%$. Similarly, the precision drops approximately $26\%$ at $MV$=$20\%$ compared with $MV$=$5\%$ when  $MR$=$20\%$. These results indicate that when $MR$ is higher, precision mainly depends on the activities of rogue RSUs. On the other hand, for lower values of $MR$, the precision values are dominated by rogue vehicles' malicious activities.

\subsubsection{Recall} 
The proposed solution identifies nearly all the malicious RSUs shown in Fig. \ref{fig:recall}. The recall value is around $92\%$ at $MV$=$20\%$, and $MR$=$60\%$. The recall ratio also indicates that the proposed solution is sensitive to the increasing $MV$. The higher values of $MV$ enable the rogue RSUs to hide their malicious properties, as discussed in Section \ref{sec:FNR}. Fig. \ref{fig:fnrate} shows that FNR is visible for higher values of $MV$. Therefore, slight increase of $FNR$ at $MV$=$20\%$ in Fig. \ref{fig:fnrate} reduces the recall values at $MV$=$20\%$ for $MR$=$40\%$, and $MR$=$60\%$ shown in Fig. \ref{fig:recall}.

\subsubsection{Accuracy} 
Figure \ref{fig:accuracy} shows the accuracy of the proposed scheme. If $MV$ is fixed, accuracy decreases with increasing $MR$. As $FP$ is nearly the same, and $FN$ is rarely visible for fixed $MV$, accuracy depends on $TP$ and $TN$. Higher values of $MR$ increase both $TP$ and $FP$, reducing $FN$ and $TN$, respectively. As a consequence, accuracy decreases with higher values of $MR$ for a specific $MV$. On the other hand, for fixed $MR$, with higher $MV$, both $FP$ and $FN$ increase, decreasing $TN$ and $TP$, respectively. Hence, the accuracy decreases gradually with increasing $MV$ for a particular $MR$. The proposed scheme achieves an accuracy of approximately $86\%$ when $MR$=$60\%$ and $MV$=$20\%$.

\subsection{Network Performance Analysis}\label{sec:net_perf}
We incorporated our proposed scheme with several VANET routing protocols such as \textit{Ad Hoc On-Demand Distance Vector (\textit{AODV})} \cite{sallam2015performance}, \textit{Optimized Link State Routing (\textit{OLSR})} \cite{chouhan2015analysis}, and  \textit{Destination Sequenced Distance Vector (\textit{DSDV})} \cite{rani2011performance} to analyze the network performance. As an entity of VANET, RSU also uses these protocols for \textit{R2R} and \textit{V2R} communication \cite{chouhan2015analysis}. We used the same simulation traffic model and parameters as discussed in Section \ref{sec:exp_setup}. To generate data packets, we used $50$ vehicles and all of the 25 RSUs as source nodes and considered all entities in the traffic model as receiver nodes. The simulation was performed for varying numbers of malicious RSUs, $MR$ ($20\%$, $40\%$ and $60\%$) and a fixed number of malicious vehicles, $MV$=$20\%$.

\subsubsection{Network Performance Metrics}
\label{sec:net_param}
To analyze the network performance we used the following metrics:

\noindent\vspace{0.2mm} \textit{1) Packet Delivery Ratio (PDR):} It is the ratio of the total number of data packets received by destination nodes to the total number of packets sent from source nodes.
\begin{equation}
\label{eqn:pdr}
    PDR = \frac{Total \; no. \; of \; packets\; received}{Total \; no. \; of \; packets \;sent}
\end{equation}

\noindent\vspace{0.2mm} \textit{2) Throughput (Tp):} It is the number of data packets transmitted successfully at a given time.
{\small
\begin{equation}
\label{eqn:th}
    Tp = \frac{Total \; No. \; of \; packets\; transmitted\; successfully}{Total\; time}
\end{equation}
}

\noindent\vspace{0.2mm} \textit{3) Average End-to-End (AE2E) Delay:} It is the ratio of the time required to send data packets from source to destination to the total number of packets received.
\begin{equation}
\label{eqn:e2e}
    AE2E\; Delay = \frac{\sum(Time\; to \;  receive-Time\; to \; sent)}{Total\; no. \; of \; packets\; received}
\end{equation}

In the subsequent sections, we analyzed the performance of the network considering the presence and absence of the proposed trust model in all the routing protocols mentioned in Section \ref{sec:net_perf}. 

\subsubsection{Packet Delivery Ratio (PDR)}\label{sec:pdr}
\begin{figure}
    \centering
    \subfloat[AODV]{\includegraphics[ width=0.23\textwidth]{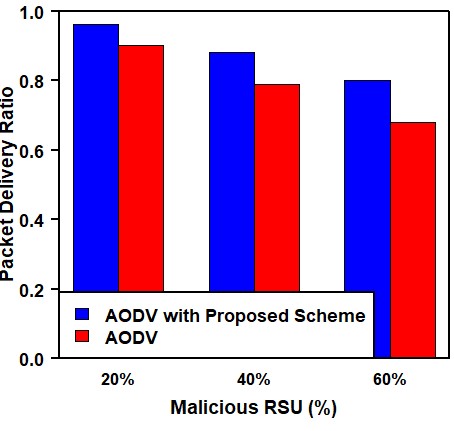}
    \label{fig:pdr_aodv}}
    \subfloat[OLSR]{\includegraphics[ width=0.23\textwidth]{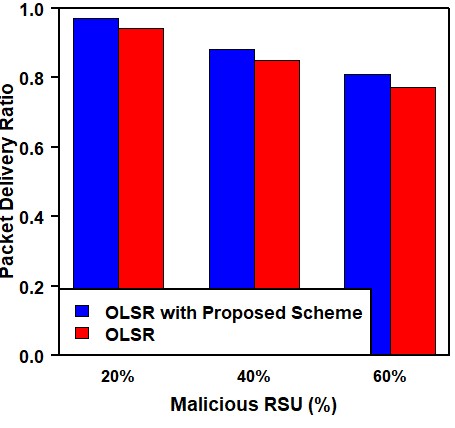}
    \label{fig:pdr_olsr}}
    \hfil
    \subfloat[DSDV]{\includegraphics[ width=0.23\textwidth]{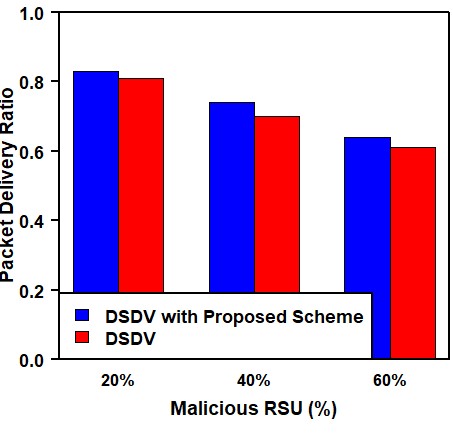}
    \label{fig:pdr_dsdv}}
    \subfloat[Protocol comparison when $MV$=$20\%$ and $MR$=$40\%$]{\includegraphics[ width=0.23\textwidth]{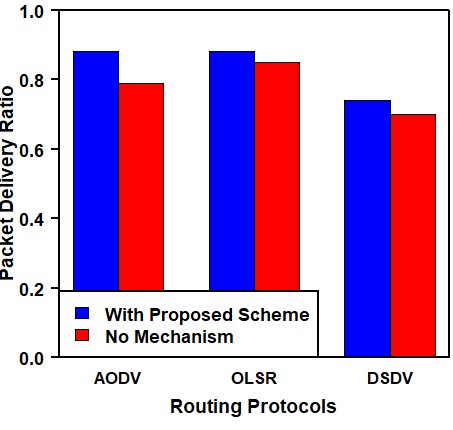}
    \label{fig:pdr_compare}}
    \caption{Packet Delivery Ratio when $MV$=$20\%$.\label{fig:pdr_all}}
\end{figure}
In the fundamental \textit{AODV}, \textit{OLSR} and \textit{DSDV} protocols where the malicious RSU detection mechanism is missing, an RSU forwards packets to next-hop RSU without considering the malicious behavior of that RSU. If the next-hop RSU is malicious, it can drop packets, resulting in low \textit{PDR}. In case of packet drops, packet re-transmissions take place in each protocol. However, integration of the proposed trust model improves the PDR in all the protocols as an RSU can decide to exclude one-hop malicious RSUs from packet forwarding. As shown in Fig. \ref{fig:pdr_aodv}, \ref{fig:pdr_olsr} and \ref{fig:pdr_dsdv}, the average improvement for \textit{AODV}, \textit{OLSR}, and \textit{DSDV} is approximately $12\%$, $4\%$, and $4\%$, respectively. Nevertheless, PDR decreases with the increasing number of malicious RSUs as they can jam the network through excessive beacon broadcasting or increase packet drop. From Fig. \ref{fig:pdr_compare} it is clear that the basic \textit{OLSR} protocol performs better than the other fundamental protocols. If any disconnection occurs, \textit{OLSR} finds a new route faster than other protocols using routing tables. In contrast, \textit{DSDV} takes a longer time to find a new route and, therefore, results in low PDR. In case of \textit{AODV}, it is not facilitated like \textit{OLSR} to get route information from some selected nodes known as MultiPoint Relay (MPR). Hence, \textit{AODV} has lower PDR compared to \textit{OLSR}. On the other hand, incorporation of the proposed trust model results in similar PDR for both \textit{AODV} and \textit{OLSR} as \textit{AODV} usually creates a route immediately if needed, whereas \textit{OLSR} updates the routing table periodically. During route discovery, \textit{AODV} also takes the advantages of the trust model to avoid malicious RSUs. Hence, PDR increases significantly for \textit{AODV}.

\begin{figure}
    \centering
    \subfloat[AODV]{\includegraphics[ width=0.5\linewidth]{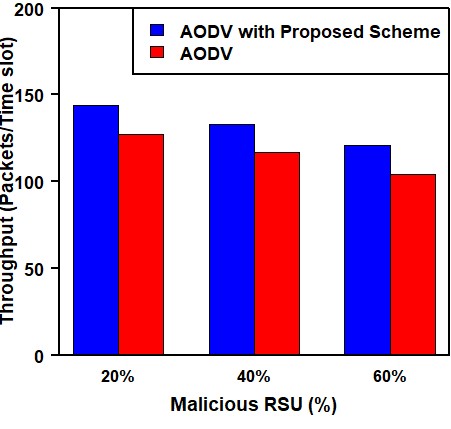}
    \label{fig:th_aodv}}
    \subfloat[OLSR]{\includegraphics[ width=0.5\linewidth]{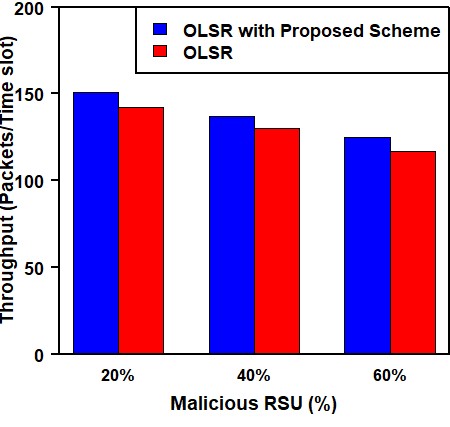}
    \label{fig:th_olsr}}
    \hfil
    \subfloat[DSDV]{\includegraphics[ width=0.5\linewidth]{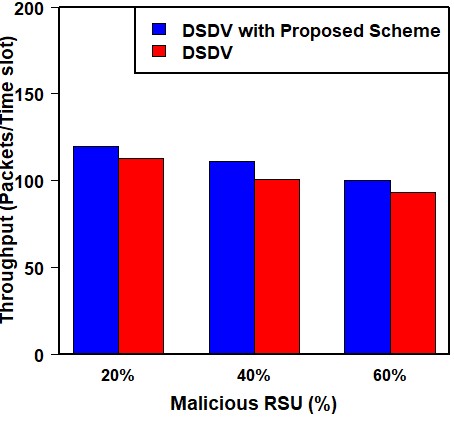}
    \label{fig:th_dsdv}}
    \subfloat[Protocol comparison when $MV$=$20\%$ and $MR$=$40\%$]{\includegraphics[ width=0.5\linewidth]{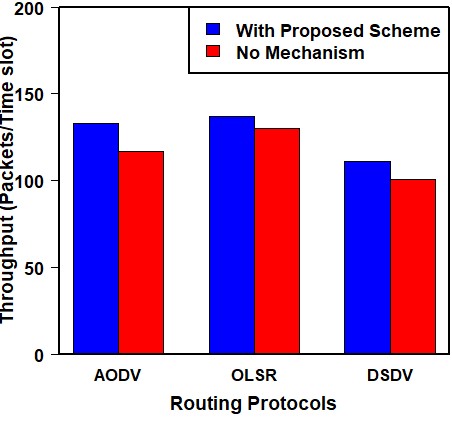}
    \label{fig:th_compare}}
    \caption{Throughput when $MV$=$20\%$.\label{fig:th_all}}
\end{figure}
\subsubsection{Throughput (Tp)}
Figure \ref{fig:th_all} presents the throughput of each routing protocol with/without the proposed trust model. As discussed earlier, each routing protocol with the trust model improves the packet delivery ratio and as a result the throughput increases. As shown in Fig. \ref{fig:th_aodv}, \ref{fig:th_olsr} and \ref{fig:th_dsdv}, the average improvement for \textit{AODV}, \textit{OLSR}, and \textit{DSDV} is approximately $14\%$, $6\%$, and, $8\%$, respectively. Similar to the \textit{PDR}, we observe from Fig. \ref{fig:th_compare} that the \textit{OLSR} protocol exhibits best throughput, which is followed by the performance of \textit{AODV} and \textit{DSDV} protocols, respectively due to their underlying mechanism as mentioned in Section \ref{sec:pdr}.

\subsubsection{Average End-to-End (AE2E) Delay}
\begin{figure}
    \centering
    \subfloat[AODV]{\includegraphics[ width=0.5\linewidth]{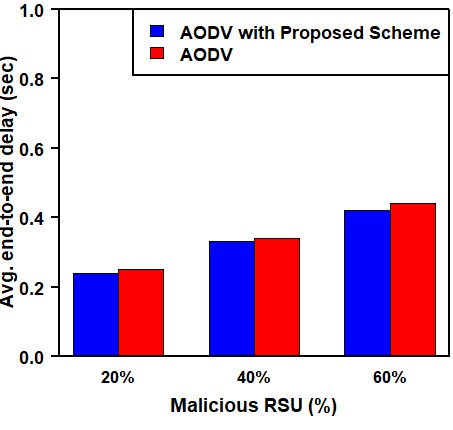}
    \label{fig:e2e_aodv}}
    \subfloat[OLSR]{\includegraphics[ width=0.5\linewidth]{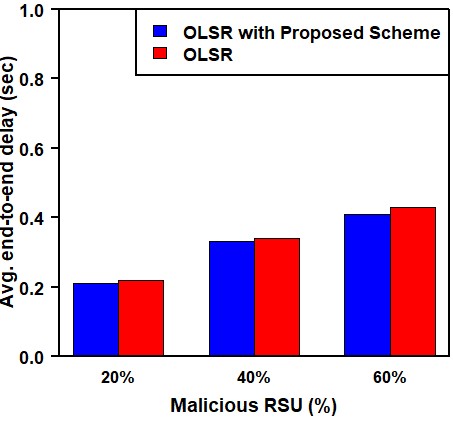}
    \label{fig:e2e_olsr}}
    \hfil
    \subfloat[DSDV]{\includegraphics[ width=0.5\linewidth]{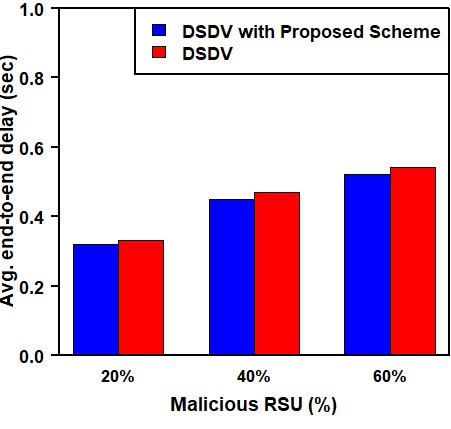}
    \label{fig:e2e_dsdv}}
    \subfloat[Protocol comparison when $MV$=$20\%$ and $MR$=$40\%$]{\includegraphics[ width=0.5\linewidth]{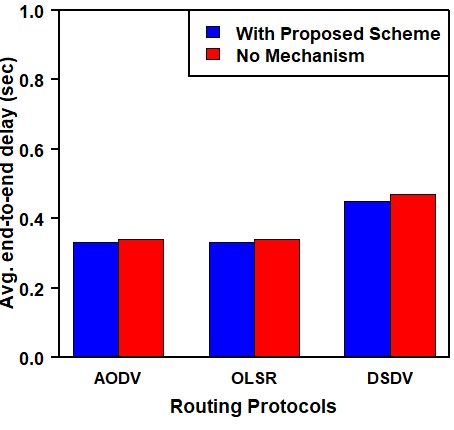}
    \label{fig:e2e_compare}}
    \caption{Average end-to-end delay when $MV$=$20\%$.\label{fig:e2e_all}}
\end{figure}

When the proposed trust model merges with the routing protocols, they select honest next-hop RSU to propagate messages. Hence, packet drops are reduced for each protocol which ultimately reduces packet retransmissions. As a consequence, end-to-end delay decreases as shown in Fig. \ref{fig:e2e_all}. However, it is observed that the improvement in the end-to-end delay is minimal. Our proposed model only detects the malicious RSUs and does not exclude them from the network. Therefore, malicious impact such as beacon flooding remains in the network that can cause congestion. Though end-to-end delay for both \textit{OLSR} and \textit{AODV} is nearly same as shown in Fig. \ref{fig:e2e_compare}, \textit{OLSR} shows slightly better performance due to the routing efficiency. Once again, \textit{DSDV} shows the worst end-to-end delay performance.
\section{Conclusion and Future Work}\label{sec:conclusion}
In this paper, we proposed a trust-based malicious RSU detection mechanism in the edge-enabled VANET. Our proposed scheme analyzes R2R communication patterns to find out deviation in RSUs' behavior and assign trust scores accordingly to distinguish malicious RSUs from nonmalicious ones. Besides, we proposed a mechanism to evaluate trust values based on the correctness of the beacon content provided by an RSU. The simulation results reveal that our scheme detects approximately $92\%$ malicious RSUs and decides the type of RSUs with an accuracy of nearly $86\%$ in the presence of rogue vehicles. Besides, the proposed scheme contributes a moderate network performance improvement of 14\% when incorporated with the \textit{AODV} routing protocol. In the future, we plan to include a sophisticated mechanism to minimize the impact of malicious vehicles. Besides, we also aim to address the IP spoofed attacks and the fuzzy behavior of VANETs in the proposed scheme. In addition, the effect of RSU-vehicle coordinated attacks on RSU trust calculation is also an interesting research direction.


\bibliographystyle{IEEEtran}
\bibliography{IEEEabrv, bibfile}
\end{document}